\documentclass[twocolumn,letterpaper,amsmath,amssymb,floatfix,aps,superscriptaddress]{revtex4}

\usepackage{graphicx}
\usepackage{dcolumn}
\usepackage{bm}
\usepackage[usenames]{color}
\usepackage{hyperref}
\usepackage{ulem}

\begin{document}

\title{Repulsive Casimir interaction: Boyer oscillators at nanoscale}%

\author{Jalal Sarabadani}
\email{jalal.sarabadani@aalto.fi}
\affiliation{Department of Applied Physics and COMP Center of Excellence, Aalto University School of Science, 
P.O. Box 11000, FI-00076 Aalto, Espoo, Finland}

\author{Badrosadat Ojaghi Dogahe}
\address{Department of Physics, University of Isfahan, Isfahan 81744, Iran}

\author{Rudolf Podgornik}
\address{Department of Physics, University of Massachusetts, Amherst, Massachusetts 01003, USA,
and Department of Theoretical Physics, J. Stefan Institute, and Department of Physics, 
Faculty of Mathematics and Physics, University of Ljubljana, SI-1000 Ljubljana, Slovenia}

\begin{abstract}
We study the effect of temperature on the repulsive Casimir interaction between an ideally permeable and 
an ideally polarizable plate {\it in vacuo}. At small separations or for low temperatures the quantum  
fluctuations of the electromagnetic 
field give the main contribution to the interaction, while at large separations or for high temperatures the 
interaction is dominated by the classical thermal fluctuations of the field. At intermediate 
separations or finite temperatures both the quantum and thermal fluctuations contribute. For a system 
composed of one infinitely permeable plate between two ideal conductors 
at a finite temperature,  we identify a {\it stable mechanical equilibrium} state, if the infinitely permeable 
plate is located in the middle of the cavity.
For small displacements the restoring force of this {\it Boyer oscillator} is linear in the deviation from the 
equilibrium position, with a spring constant that depends inversely on the separation 
between the two conducting plates and linearly on temperature. Furthermore, an array of such 
oscillators presents an ideal Einsteinian crystal that displays a fluctuation force between its outer 
boundaries stemming from the displacement fluctuations of the Boyer oscillators.
\end{abstract}

\maketitle

A long-range attraction between two ideal flat conductors due to fluctuations of the electromagnetic 
(EM) field at zero temperature has been discovered by H.B.G. Casimir in 1948 \cite{Casimir} and is 
directly connected with the change in the quantum vacuum zero-point energy~\cite{milonnibook}. 
The magnitude of the Casimir attractive force per unit surface area is,  
$\frac{F^{{C}} }{A} = -\frac{\hbar c \pi^2}{240 H^4}$, where $A$ is the surface area of the plates, 
$\hbar$ is the Planck constant divided by $2\pi$, $c$ is the speed of light in the vacuum and $H$ is the 
separation between two plates.  In 1955 Lifshitz 
developed a more general framework to investigate forces between two dispersive 
dielectric media at finite temperature $T$ 
\cite{Lifshitz}. The ensuing {\it Casimir-Lifshitz interactions} have gained much attention  
\cite{ParsegianBook,french,Bordagbook,kardarRMP1999,jalalPRA2006, jalalPRA2007,soltaniPRA2010a,soltaniPRA2010b,%
soltaniAnnals2011,jalalPRA2011,jalalPRB2011,zakeriPRA2012} 
as they are one of the direct macroscopic manifestations of the quantum theory.  {Investigating the 
Casimir interaction with asymmetric boundary conditions, Boyer in 1974 showed that an infinitely polarizable (ideal 
conductor) and an infinitely permeable plate (a  {two-plate} Boyer setup) at $T=0$K repel each other with a repulsive force 
$F=-(7/8) F^{{C}}$ \cite{boyer1974}. Later, this result was confirmed by two different methods, the 
radiation pressure method \cite{hushwater} and the path-integral formalism \cite{kiani2012}. 
Boyer's conclusions are in general consistent with the Lifshitz theory where repulsive Casimir interactions 
are only possible in asymmetric interaction setups \cite{ParsegianBook}, such as a surface with high 
dielectric response (infinite polarizability of an ideal conductor) apposed to a surface with high magnetic 
response (infinite permeability) across vacuum \cite{Woods}. }

\begin{figure}[b]
\begin{center}
  \includegraphics[width=0.3\textwidth]{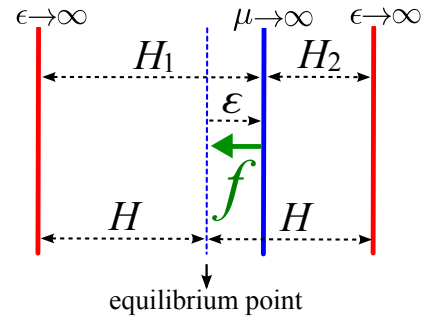}
\caption{Schematic depiction of the Boyer oscillator. Two parallel flat ideal conductors ($\epsilon \rightarrow \infty$)
apposed at separation $2H$,  with a parallel flat ideally permeable ($\mu \rightarrow \infty$) plate 
between them. When the permeable plate is moved to the right for $\varepsilon$, which means that 
$H_1 = H +\varepsilon$ and $H_2 = H -\varepsilon$, the Casimir force $F$ acts on the permeable plate to bring it back to its 
equilibrium point.}
\label{fig:schematic_nano_oscillator}
\end{center}
\end{figure}

Repulsive Casimir-Lifshitz interactions are well documented also experimentally 
\cite{Repulsion1,Repulsion2,Repulsion3}.  The Casimir-Lifshitz 
force between a gold sphere ($\epsilon_G$)  and a silica substrate ($\epsilon_S$), when they are 
immersed in bromobanzene ($\epsilon_B$), has been found to be repulsive ($\epsilon_G> \epsilon_B 
> \epsilon_S$)  \cite{ParsegianNature2009}, just like the force between a permeable yttrium 
iron garnet (YIG) plate and a ferroelectric BaTiO$_3$ {\it in vacuo}, possibly implicating the Boyer mechanism
in the latter case \cite{Shao}. The {calculated} Boyer force per unit 
area then follows as $\sim 1.3 \times 10^{-4} {\textrm N}/{\textrm m}^2$ if the plates are
separated by $1 \mu\textrm{m}$, becoming significantly larger at sub-micron scales.
Casimir-Lifshitz interactions are particularly strong on the nanoscale \cite{french} and 
must therefore be taken into account when designing micro- and nano-electromechanical systems 
(MEMS and NEMS) \cite{MEMSdesigne1,MEMSdesigne2}.

{In order to unravel additional details of the repulsive Casimir-Lifshitz interaction we analyze 
a Boyer setup composed of two ideally polarizable ($\epsilon 
\rightarrow \infty$) plates with an ideally permeable ($\mu \rightarrow \infty$) plate in the middle, 
see Fig. \ref{fig:schematic_nano_oscillator}, at 
separation $H$ and temperature $T$. We explore the possibility of stable equilibria and construct a 
mechanical oscillator that furthermore leads to an interesting 1D ideal Einsteinian crystal based on 
the Boyer repulsive interaction.}

We start with the two-plate Boyer setup. A point on each 
plate is described by $x_{\alpha}(x)=(\mathbf{x}, \delta_{\alpha,2} H)$,  where  $\alpha= 1$ and $2$ identifies the 
plate, $\delta_{\alpha,2}$ is the Kronecker delta function, $x=(\mathbf{x},x_{0})$,  $x_0$ is the temporal component, 
and $\mathbf{x}=(x_{1}, x_{2})$ defines two lateral spatial coordinates.
For the geometry we are considering, by decomposing the EM field into the transverse magnetic (TM) and 
the transverse electric (TE) waves \cite{jacksonBook} all components of the EM field can be expressed by
the scalar fields $\Phi_{\textrm{TM}} = B_{\parallel}$ and $\Phi_{\textrm{TE}} = E_{\parallel}$, 
where $B_{\parallel}$ and $E_{\parallel}$ are the components of the magnetic and electric 
fields parallel to the surfaces of the plates. For the scalar field $\Phi_{\textrm{TM}}$ Dirichlet (D)
and Neumann (N) boundary conditions (BCs) are satisfied by the infinitely polarizable and infinitely permeable
plates, respectively, while for the scalar field $\Phi_{\textrm{TE}}$  an N BC and a D BC are satisfied 
by the infinitely polarizable and infinitely permeable plates, respectively \cite{kiani2012}.
Using the Matsubara formalism \cite{kapustabook,jalalPRA2011} the partition function 
in the presence of two parallel plates can be written as 
\begin{equation}
Z= \int \big[ {\mathcal{D}} \Phi \big]_{\mathcal{C}}e^{-S[\Phi]/\hbar},
\end{equation}
where we ignored the inessential normalization w.r.t 
the partition function of the free space. $\Phi$ is a scalar field, subscript $\mathcal{C}$ denotes the  constraint 
imposed by the plates on the scalar field and the model Euclidean action describing the field is assumed to be
\begin{equation}
S[\Phi]= \frac{1}{2}\int_{0}^{\beta}\!\!\!d \tau\!\!\int\!\!d^{3}x\left[\left( \frac{\partial \Phi}{\partial \tau }   
\right)^{2}+(\nabla\Phi)^{2}\right].
\end{equation}
Here $\beta= \frac{1}{k_B T}$, with $k_B$ the Boltzmann constant. As the scalar bosonic field satisfies the periodicity 
condition, $\Phi(x,\beta)=\Phi(x,0)$, one can expand it as $\Phi (x, \tau)= \sum_{n=-\infty}^{\infty}  \Phi_n (x) 
e^{i \omega_n \tau}$  where $\omega_n = \frac{2 \pi n k_B T}{\hbar}$ are the Matsubara frequencies. 
$\Phi(x,\tau)$ is real and must satisfy mixed boundary conditions, as explained above \cite{kiani2012}, 
i.e. {for TM waves} a  D BC must be satisfied at the infinitely polarizable plate, while an N BC 
must be satisfied at the infinitely permeable plate. The Matsubara  approach \cite{kapustabook,jalalPRA2011} 
then yields the partition function for a pair of plates, where one has an N and the other one a D BC as 
(up to a multiplicative constant)
\begin{equation}
Z_{D N}=
\int\!\!\!\prod_{n=- \infty}^{   \infty}\!\!\!{\mathcal{D}} \Phi_n \delta (\Phi_n (X_1)) 
\delta (\partial_{N} \Phi_n (X_2))e^{-S[\Phi]}
\end{equation}
where $\partial_{N}$ is the partial  derivative in the normal direction to the surface. Integrating over 
$\Phi_n$ the logarithm of the partition function  {for the TM waves} can be expressed as 
$\ln Z_{D N} = \sum_{n= -\infty}^{\infty} \ln \{ det [\Gamma_n^{D N}] \}$ where 
\begin{equation}
\Gamma_n^{D N} =\left( 
\begin{array}{cc}
G_n(x-y, 0) & \partial^{\prime}_{z} G_n(x-y, -H) \\ 
\partial_{z} G_n(x-y, H) & \partial_{z}\partial^{\prime}_{z} G_n(x-y, 0),
\end{array}
\right) 
\end{equation}
with the free space Green's function $G_n(X _{ \alpha } (x),Y _{ \sigma }(y))= 
\frac{e ^{- \omega _{n} \vert X _{ \alpha } (x)-Y _{ \sigma }(y) \vert } }
{4\pi \vert  X _{ \alpha } (x)-Y _{ \sigma }(y) \vert } $. The prime is used for the derivative with respect 
to the second variable, i.e. $y$. 
{A similar procedure can be applied for the TE waves to obtain the logarithm of the partition function,
$\ln Z_{N D}$, 
with an N BC on the infinitely polarizable plate and a D BC on the infinitely permeable plate, and 
$\ln Z_{N D}= \ln Z_{D N}$.
}
\begin{figure}[t]
\begin{center}
  \includegraphics[width=0.48\textwidth]{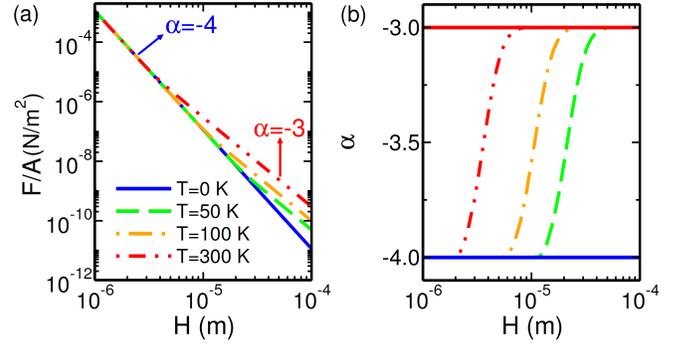}
\caption{(a) Boyer force per unit area, $F(T, H)/A$, as a function of the separation between two plates, $H$,
for various values of temperature $T=0-300$K (from bottom to top). $\alpha$ is the force exponent which 
is defined as $F/A \sim H^{\alpha}$. $\alpha= -4$ and $-3$ identify the quantum and classical regimes, 
respectively. (b) $\alpha$ as a function of $H$ for various values of temperature as of panel (a). Here 
blue and red horizontal solid lines pertain to the quantum and classical regimes, respectively.}
\label{fig:exponent_force}
\end{center}
\end{figure}
The normal Boyer force between the two plates can then be obtained from the free energy 
${\cal F} = - k_BT {[} \ln Z_{N D}  {+ \ln Z_{D N} ]}$ as 
$F = - \frac{\partial {\cal F}}{\partial H}$, with 
\begin{equation}
\frac{ F(T, H)}{A} = k_B T  \sum_{n=- \infty}^{ \infty} \int_0^{\infty} \frac{p dp}{\pi }   
\frac{ \rho_n  }{1+e ^{2\rho_n H }},
\label{Boyer_force_two_plates}
\end{equation}
where $$\rho_n = \sqrt{p ^2+ \omega_{n}^{2} /c^2 },$$ 
$p^2 = p _1^{2} + p_2^2$, and $p_1$ and $p_2$ correspond to $x_1$ and $x_2$ coordinates in the 
Fourier space, respectively.
The force at $T = 0$K can be calculated by replacing the Matsubara summation with a continuous integral as  
$ \frac{F(T=0{\textrm{K}},H)}{A}=\frac{7}{8} \frac{ \hbar c \pi ^{2}}{240H ^{4} }   = -\frac{7}{8} F^{{C}}$. This expression is 
in complete agreement with the previous result of Boyer \cite{boyer1974,hushwater,kiani2012}. 
At finite temperature the force can be cast into an alternative form
\begin{eqnarray}
 \frac{F(T, H)}{A}&=&\frac{3 k_B T \zeta(3) }{16\pi  H ^{3}  } + 
\frac{2 k_B T}{\pi}\!\sum_{n=1}^{ \infty}\!\int_{0}^{ \infty}\!\!\!\!\!\! p dp \frac{ \rho_n }
{1 + e ^{2 \rho_n H }}  \nonumber\\
&=&k_B T  \sum_{n=1}^{ \infty} F_n(T, H) ,
\label{Eq:Casimir_force_two_flat_plates}
\end{eqnarray}
where the first term, $\frac{3 k_B T \zeta(3) }{16\pi H ^{3}  }$, is the zero Matsubara frequency term which is the 
value of  the force at large separations or high temperatures. Comparing this with the Casimir interaction force 
${F^C(T, H)}$ between two ideally polarizable (metallic) sheets at finite temperature 
\cite{ParsegianBook,Bordagbook,jalalPRA2011,jalalPRB2011}, 
we note that
\begin{equation}
F_{n}(T, H) = - F_{n}^C(T, H) + 2 F_{n}^C(T, 2H),
\end{equation}
so that in every regime the Boyer interactions for the asymmetric ideally permeable - ideally conductive system 
can be mapped onto a symmetric Casimir interaction for ideal conductors.

In Fig. \ref{fig:exponent_force}a we plot the Boyer force per unit area as a function of $H$ for various values of 
the temperature $T=0-300$K. For separations smaller than the thermal length 
$\lambda_T = \hbar c/ (2 \pi k_B T) \sim 1.2$$\mu$m at room  temperature ($T= 300$K), all curves collapse onto 
the zero temperature curve, revealing that the quantum fluctuations play the main role in the interaction, with 
the force scaling as $H^\alpha$ and $\alpha = -4$. For $H > 1.2 \mu$m, as the temperature 
increases, the role of thermal fluctuations is gaining in importance and at large separations it is the thermal 
fluctuations that dominate the interaction with the force exponent $\alpha = -3$. To distinguish the 
quantum from the classical regimes, Fig. \ref{fig:exponent_force}b shows the force exponent, $\alpha$ as 
a function of $H$. The transition between the quantum and the classical regimes 
is shifted to smaller separations for higher temperatures. For comparison, the attractive force
between two ideal metallic sheets goes from $\alpha = -4$ to $\alpha = -3$ on increase of the separation 
\cite{jalalPRA2011,jalalPRB2011}.

\begin{figure*}[t]
\begin{center}
  \includegraphics[width=0.98\textwidth]{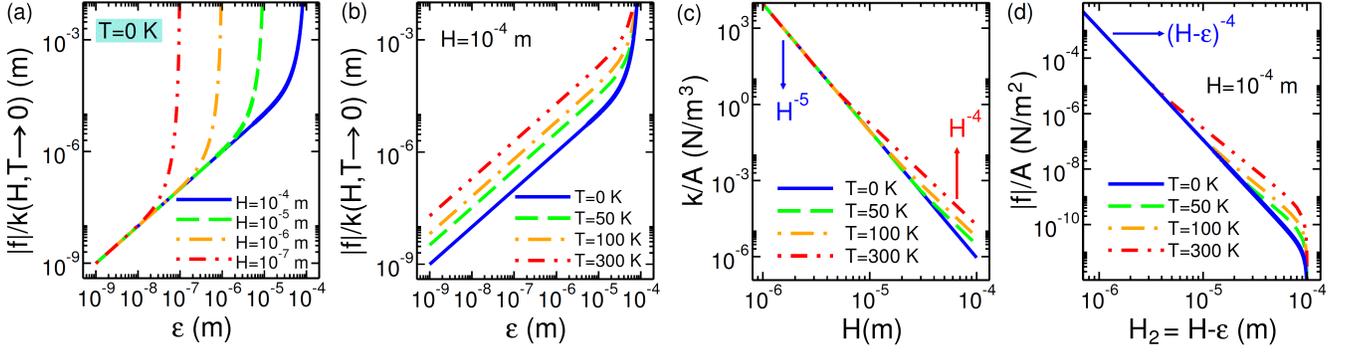}
\caption{(a) The magnitude of the normalized Boyer force, $|f| / k(H, T \rightarrow 0)$, as a function of the 
displacement of the middle ideally permeable plate from its equilibrium position, $\varepsilon$, for various values of 
the separation between equilibrium position of the permeable plate and each conductor, $H=10^{-7}-10^{-4}$m 
(from top to bottom). Here $k(H, T \rightarrow 0)$ is the spring constant of the Boyer interaction at zero temperature. 
(b) $|f| / k(H, T \rightarrow 0)$  at constant $H=10^{-4}$m has been plotted as a function of $\varepsilon$ for various 
values of temperature $T=0-300$K (from bottom to top). (c) Spring constant per unit area, $k/A$, as a function of $H$ 
for various values of temperature  as of panel (b). (d) The magnitude of the force per unit area $|f|/A$, at constant 
$H=10^{-4}$m as a function of  the separation between permeable plate and right conductor, $H_2=H-\varepsilon$, for 
various values of $T$ as of  panels (b) and (c).}
\label{fig:force_nano_oscillator}
\end{center}
\end{figure*}

In the double Boyer setup composed of three flat parallel plates, one ideally permeable between two ideally 
polarizable ones, all immersed in vacuo at finite temperature $T$ (see Fig. \ref{fig:schematic_nano_oscillator}), 
the Matsubara  approach \cite{kapustabook,jalalPRA2011} or the transfer matrix formalism 
\cite{rudiJCP2003,jalalPRB2011,rudiJCP2006} both yield the free energy per unit area as 
\begin{equation}
\frac{ {\cal F}^s}{A} = k_B T \sum_{n=-\infty}^{\infty}  {\cal F}_n^s (H_1,H_2),
\label{Eq:Casimir_Free_Energy} 
\end{equation}
where 
\begin{equation}
{\cal F}_n^s (H_1,H_2)\!= \!\!\int_{0}^{\infty}\!\frac{p dp}{2\pi} 
\ln \! \big[ 1 + e_n(H_1, H_2) \big],
\end{equation}
and $$e_n (H_1, H_2) = e^{{-}2\rho_n H_1} \!+\! e^{{-}2\rho_n H_2} \!+\! 
e^{{-}2\rho_n (H_1+H_2)},$$
with $H_1$ and $H_2$ the separations between left 
and right hand conductors and the middle permeable plate, thus implying $H_1 + H_2 = 2H$ as shown in 
Fig. \ref{fig:schematic_nano_oscillator}. 
The force $f$ per unite area that acts on the middle permeable plate
can be calculated as 
\begin{equation}
\frac{f}{A}= - \frac{\partial }{\partial H_1}\!\!\left(\frac{{\cal F}^s}{A}\right) 
+\frac{\partial }{\partial H_2}\!\!\left(\frac{{\cal F}^s}{A}\right) =  
k_B T \!\!\!\sum_{n=-\infty}^{\infty} f_n (H_1,H_2),
\label{Eq:Casimir_Force}
\end{equation}
where $$f_n (H_1,H_2)\!=\!\!\int_{0}^{\infty}\! \frac{p dp }{\pi} 
\frac{ \rho_n [ e^{-2\rho_n H_1} - e^{-2\rho_n H_2} ] }{1 + e_n(H_1, H_2)}.$$ 
The total force on the middle plate is thus a superposition of the two forces that act from the left and 
from the right conductors on the middle, ideally permeable plate. As these two forces are repulsive, it follows 
from the symmetry of the system that the {\it stable equilibrium} corresponds to $H_1=H_2$, see 
Eq. (\ref{Eq:Casimir_Force}). This is not at odds with the general theorem on the absence of stable equilibria 
with Casimir forces, since it hinges upon interactions only between objects within the same class of material  
\cite{Rahi}, an assumption obviously violated for the Boyer setup.

When the permeable plate is displaced towards either of the conductor plates by $\varepsilon$, the 
restoring force that acts on it can be calculated as a function of $\varepsilon$ by setting $H_1 = 
H +\varepsilon$ and $H_2 = H -\varepsilon$.  The direction of the force is always toward the stable 
equilibrium location (see Fig. \ref{fig:schematic_nano_oscillator}) and its magnitude can be obtained to the first 
order w.r.t. $\varepsilon$ in the Hookian form, $f = - k \varepsilon$, with the spring constant $k = k(H, T)$ defined as 
\begin{equation}
\frac{k(H, T)}{A} = k_B T \sum_{n=-\infty}^{\infty} k_n (H,T),
\label{Eq:Spring_Constant}
\end{equation}
where $$k_n (H, T) = \int_{0}^{\infty} \frac{p dp}{\pi} \frac{ 4 \rho_n^2 e^{-2 \rho_n H}}{ [1 + e^{-2 \rho_n H}]^2}.$$ 
At low temperatures or small separations (quantum limit) the spring constant per unit area is 
$\frac{k(H, T \rightarrow 0)}{A} \rightarrow \frac{7\hbar c\pi^2}{240 H^5}$, while at high temperatures or large 
separations (classical limit)  $\frac{k(H, T \rightarrow \infty)}{A} \rightarrow \frac{9 \zeta (3) k_B T }{8 \pi  H^4}$. 
In Fig. \ref{fig:force_nano_oscillator}a we plot the magnitude 
of the normalized  force, $|f|/k(H, T \rightarrow 0)$, as a function of $\varepsilon$, for various values of 
$H=10^{-7}-10^{-4}$m (from top to bottom). 
As is evident from the plot, the Hookian limit is valid within the range $|\varepsilon| < H/10$. In 
Fig. \ref{fig:force_nano_oscillator}b the magnitude of the normalized Casimir force 
has been plotted as a function of $\varepsilon$ at constant $H=10^{-4}$m for $T=0-300$K 
(from bottom to top). By increasing the temperature the spring constant also increases with 
the range of validity of the Hookian region remaining unchanged.  

Fig. \ref{fig:force_nano_oscillator}c presents the 
spring constant, $k(H, T)$, per unit area as a function of $H$ for various values of temperature $T=0-300$K 
(from bottom to top). Obviously at low temperatures or small separations the spring constant 
per unit area scales as $H^{-5}$, while at high temperature or large separations it scales 
as $H^{-4}$. Fig. \ref{fig:force_nano_oscillator}d shows $|f|/A$ at $H=10^{-4}$m as a function 
of $H_2=H-\varepsilon$ for various values of temperature $T=0-300$K (from bottom to top). 
For small values of $H_2$ the contribution of quantum fluctuations to the interaction dominates 
over that of the classical thermal fluctuations. Indeed at sufficiently 
small $H_2$ the force per  unit area is $f/A = \frac{7}{8} \frac{\hbar c \pi^2}{240 (H-\varepsilon)^4}$. 

If the middle permeable plate is free to move, the eigenfrequencies of such Boyer oscillator 
at a finite temperature are then given by
$\omega^2(H, T) =  {\frac{k(H, T)}{m}}$ where $m$ is the mass of the plate. 
Assuming it has a thickness $d$ and a cross section area $A$, using $m= \varrho d A$, where $\varrho$ 
is the mass density, the eigenfrequency in the two temperature limits becomes 
\begin{equation}
\omega^2(H, T \rightarrow 0)= {\frac{7 \hbar c \pi^2}{240 H^5 \varrho d}},
\label{eghrwk1}
\end{equation} 
and 
\begin{equation}
\omega^2(H, T \rightarrow \infty)= {\frac{9 \zeta(3) k_BT}{8 \pi H^4 \varrho d}}. 
\label{eghrwk2}
\end{equation}
Interestingly, $\omega$ is not a function of the cross section area $A$. Given the 
mass density of YIG as $\varrho = 5.172 $gr/cm$^3$ 
the magnitude of the frequency for $d=12.4$nm is $\omega(H, T \rightarrow 0)=1.9$kHz, and $0.6$MHz at 
$H=1\mu$m and $H=0.1\mu$m, respectively, and 
$\omega(H, T \rightarrow 0)=0.6$GHz when $d=1.24$nm and $H=10$nm.
{In the classical regime, by choosing $H=10\mu$m and $T=300$K, the eigenfrequencies are obtained as
$\omega(H, T =300K)= 152.1$Hz, $48.1$Hz and $15.2$Hz for $d=1.24$nm, $d=12.4$nm and $d=124$nm, respectively.}

Finally, we examine the elastic modulus of the system at constant temperature,  {$K$}, 
defined as the inverse of the compressibility at constant temperature, 
$-\frac{1}{V}\big( \frac{\partial V}{\partial P} \big)_T $ where $V$ is the  volume 
and $P$ is the pressure of the system. The elastic modulus for the double Boyer setup, 
{Fig. \ref{fig:schematic_nano_oscillator}, for $H_1 =H_2$} then 
becomes
\begin{equation}
K(H, T) = \frac{H}{A} \left(\frac{\partial^2  {\cal F}^s(H, T)}{\partial H^2}\right) = \frac{H}{2 A}  {k(H, T)}.
\label{compressibility1}
\end{equation}
At very low temperatures or small separations 
${K} (H, T \rightarrow 0) \rightarrow  \frac{7 \hbar c \pi^2}{480 H^4}$  while at high 
temperature or large separation limits 
${K} (H, T \rightarrow \infty) \rightarrow \frac{ 9 \zeta (3) k_B T }{16 \pi H^3}$. 
The magnitude of the elastic modulus  for $T=0$K 
at $H= 1$nm is ${K}  =4.55$GPa which is about twice that of water at $T= 293$K. 

\begin{figure}[b]
\begin{center}
  \includegraphics[width=0.35\textwidth]{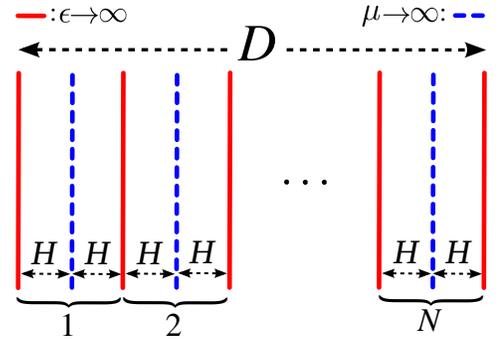}
\caption{Schematic geometrical picture of the multi-layers system composed of $N+1$ flat ideal 
conductors ($\epsilon \rightarrow \infty$) and $N$ flat ideal permeable ($\mu \rightarrow \infty$) plates. 
Each permeable plate is located between two conductors, and each conductor is located between 
two permeable plates except the first and the last conductors. The distance between
each plate with its neighboring plates is $H$, {and the full width of the Boyer oscillator array is $D= 2 N H$
in the limit of $H \gg d$, where $d$ is the thickness of each ideal plate.} }
\label{fig:schematic_multilayers}
\end{center}
\end{figure}

Interestingly, the bulk modulus for a multi-layer Boyer system ($B_N$) composed of 
$N+1$ flat ideal conductors and $N$ 
flat ideal permeable plates, depicted in  Fig.~\ref{fig:schematic_multilayers}, 
so that each permeable plate is located between two adjacent ideal conductors, with the separation between each 
plate with its neighboring plates being $H$, is the same as for the system composed of one permeable plate between 
two conductors. This is due to the fact that the conducting plates ideally screen the field.
Allowing for the free displacement of the ideally permeable plates, we then get a Boyer oscillator array,
which can in fact be described as an ideal {\it Einsteinian crystal} since the displacements of the 
different oscillators are not coupled.

The free energy of such an Einsteinian crystal {, ${\cal F}^{\textrm{EC}}$,} can then be calculated as
\begin{eqnarray}
&&\kern-30pt {\cal F}^{\textrm{EC}} (D, T) = - N k_BT \log{{\cal Z}\left({\frac{D}{2N}},T\right)} = \nonumber\\
&&\kern-30pt= \frac{N}{2}\hbar \omega\left(\frac{D}{2N}, T\right) +
{N} k_BT \log{\left(1 - e^{- \beta \hbar \omega(\frac{D}{2N}, T)}\right)},
\end{eqnarray}
where $H = {\frac{D}{2N}}$ (in the limit of $H \gg d$) and $D$ is the total width of the 
Boyer oscillator array, while ${\cal Z}({\frac{D}{2N}},T)$ is the partition function of a single Boyer harmonic oscillator.
This gives us another fluctuation force acting between the boundaries of the array, dependent on the 
original Boyer repulsion within a single oscillator in the array. In the different temperature limits we obtain from 
Eqs. (\ref{eghrwk1}) and (\ref{eghrwk2}) to the lowest order
\begin{equation}
{\cal F}^{\textrm{EC}} (D, T \rightarrow 0) = 
\frac{N^{7/2}}{2}\sqrt{\frac{7~ 2^5 ~\hbar^3 c \pi^2 }{240 ~D^5 \varrho d}},
\end{equation}
and 
\begin{equation}
{\cal F}^{\textrm{EC}} (D, T \rightarrow \infty) =  
\frac{{N} k_BT}{2} \log{\left( \frac{18 N^4}{\pi~ D^4 \varrho d}\right)}.
\end{equation}
\begin{figure}[t]
\begin{center}
  \includegraphics[width=0.4\textwidth]{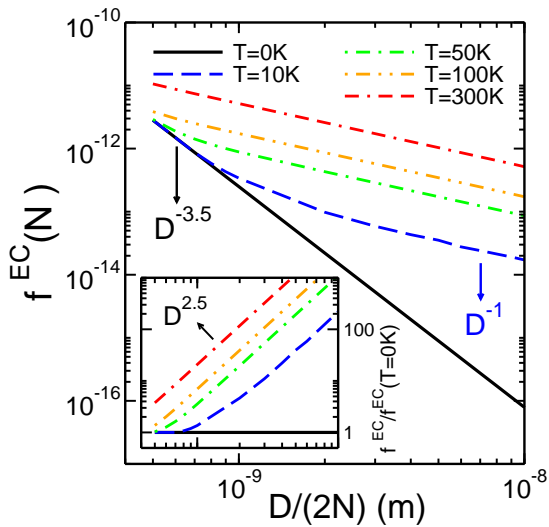}
\caption{Fluctuation force, $f^{\textrm{EC}}$, as a function of the width of each cell, $\frac{D}{2N}$, in the oscillators array,
for various values of the temperature of the system, $T=0-300$K (from bottom to top). Inset shows the normalized
fluctuation force, $f^{\textrm{EC}} / f^{\textrm{EC}} (T=0{\textrm{K}})$, as a function of $\frac{D}{2 N}$.
}
\label{fig:fluctuation_force}
\end{center}
\end{figure}
We see that in both limits the total free energy corresponds to a repulsive interaction and is  
not extensive in the number of oscillators, while the 
dependence on the width of the Boyer oscillator array has a different form than the $H$ dependence of the
Boyer interaction for each of the oscillators. Obviously for large temperatures the fluctuation interaction 
between the boundaries of  
a Boyer array is an anomalously long-ranged function of the extension of the array, decaying algebraically as $D^{-1}$. 
The fluctuation force between the boundaries of the Boyer oscillator
array, $f^{\textrm{EC}} = - \frac{\partial {\cal F}^{\textrm{EC}} }{\partial D}$, can be derived as
\begin{equation}
f^{\textrm{EC}} (D,T \rightarrow 0 ) = 5 \bigg(\frac{N}{D} \bigg)^{\frac{7}{2}} ~ \sqrt{\frac{7 \hbar^3 c \pi^2}{120 \varrho d}},
\label{fluctuation_force_zero-Temp}
\end{equation}
and 
\begin{equation}
f^{\textrm{EC}} (D,T \rightarrow \infty ) = \frac{2N k_B T}{D},
\label{fluctuation_force_high-Temp}
\end{equation}
respectively.
In Fig. \ref{fig:fluctuation_force} the fluctuation force, $f^{\textrm{EC}}$, has been plotted as a function of the width of 
each cell, $\frac{D}{2N}$, in the oscillator array, for various values of the temperature of the system, $T=0-300$K 
(from bottom to top), with $d=1.24$\AA ~and $\varrho=5.127{\textrm{gr/cm}}^3$. As is clear, for low temperatures and 
small distances limits the force scales as $D^{-\frac{7}{2}}$, while for
large separations or high temperatures it scales as $D^{-1}$. To show these scaling behaviors more clearly the normalized 
fluctuation force, $f^{\textrm{EC}} / f^{\textrm{EC}} (T=0{\textrm{K}})$, has been plotted as a function of $\frac{D}{2 N}$. 
The slope of the curve for the normalized force at $T=300$K is equal to +2.5 for the whole range of the separations which 
reveals the high temperature limit. As the temperature decreases the slope of the normalized force curve decreases at small 
separations due to the quantum fluctuations.

Boyer interaction between a flat ideal conductor and a flat ideally permeable plate is repulsive and 
depends strongly on  the temperature of the system. 
At small separations or very low temperatures, only quantum fluctuations play a role 
in the interaction (quantum regime), while at intermediate  separations or moderate temperatures (transient regime) 
both quantum and thermal fluctuations are important. At large  separations or high temperatures (classical limit) 
the interaction is governed mainly by the contribution of the classical thermal fluctuations. For a Boyer 
system composed 
of three flat parallel plates, one permeable plate between two conductors, all immersed in vacuo at a finite 
temperature, the system possesses a stable equilibrium when the middle permeable plate is located exactly in the 
middle of the cavity.
For small values of displacement from this stable equilibrium the force is Hookian and the corresponding spring 
constant is a function of the separation between the two bounding conductors and the temperature. Harmonic vibrations 
around this stable equilibrium value define a Boyer oscillator with temperature dependent eigenfrequencies. 
A linear array of such oscillator behaves as an ideal Einsteinian crystal and the confined
fluctuations of ideally permeable plates around their equilibrium positions induces a fluctuation interaction between
its boundaries. While being related to the Boyer interaction, producing a restoring force within a single 
oscillator, the fluctuation interaction between the boundaries of the Boyer array has a very different 
separation dependence in the limit of small and large temperatures.

\begin{acknowledgments}
J.S. acknowledges support from the Academy of Finland through its Centers
of Excellence Program (2012-2017) under Project No. 915804. R.P. acknowledges support from 
the ARRS through Grant No. P1-0055.
\end{acknowledgments}

\end{document}